**Layer-Number-Controlled Symmetry Breaking and Surface-State Transport in Rhombohedral Graphene Multilayers**

Bosai Lyu[1,2†], Jian Zheng[1†], Kai Liu[1†], Yulu Ren[1], Size Wu[1], Yating Sha[1], Shuhan Liu[1], Youngju Park[3], Kenji Watanabe[4], Takashi Taniguchi[5], Jinfeng Jia[1], Zhiwen Shi[1], Jeil Jung[3,6], Weidong Luo[1], Guorui Chen[1*]

[1] State Key Laboratory of Micro-nano Engineering Science, Key Laboratory of Artificial Structures and Quantum Control (Ministry of Education), Tsung-Dao Lee Institute and School of Physics and Astronomy, Shanghai Jiao Tong University, Shanghai, People's Republic of China.

[2] Department of Quantum Matter Physics, University of Geneva, 24 Quai Ernest Ansermet, CH 1211 Geneva, Switzerland.

[3] Department of Physics, University of Seoul, Seoul, Korea.

[4] Research Center for Electronic and Optical Materials, National Institute for Materials Science, 1-1 Namiki, Tsukuba 305-0044, Japan.

[5] Research Center for Materials Nanoarchitectonics, National Institute for Materials Science, 1-1 Namiki, Tsukuba 305-0044, Japan.

[6] Department of Smart Cities, University of Seoul, Seoul, Korea.

[†]These authors contributed equally to this work.

[*]Contact author: chenguorui@sjtu.edu.cn

## Abstract:

**Rhombohedral multilayer graphene hosts layer-polarized flat bands, providing an intriguing platform for correlated and topological electronic states; however, the role of layer number in governing symmetry breaking and surface screening remains elusive. Here we prepare rhombohedral graphene multilayers and systematically conduct electrical transport measurements. We uncover an unconventional layer dependence of phase transitions: the critical displacement field ($D_c$) for the layer-antiferromagnetic (LAF)-to-semimetal transitions remains constant across tetralayer to hexalayer graphene, whereas the $D_c$ for semimetal-to-layer-polarized-insulator (LPI) transition increases with layer number, defying unscreened Coulomb interaction models. In hexalayer graphene, surface-state-dominated transport emerges, with Landau levels (LLs) and resistive peaks selectively controlled by adjacent gates, a signature of strong interlayer screening absent in thinner stacks. High magnetic fields reveal valley-layer-locked LLs and dissipative states possibly from interlayer backscattering, highlighting the presence of decoupled surface states. Our findings establish layer number as a key tuning knob for engineering**

**correlated and topological phases in rhombohedral graphene multilayers.**

**Article Text:**

## I. INTRODUCTION

Rhombohedral multilayer graphene exhibits nearly flat bands localized at the top and bottom surfaces, with an energy dispersion relation, $E_k \propto p^N$, where $E_k$ is the kinetic energy, $p$ is the momentum and $N$ is the number of layers[1-3]. This special energy spectrum gives rise to strong $e$-$e$ interactions dominating over kinetic energy at low carrier densities, leading to many intriguing correlated and topological electronic states, such as layer-antiferromagnetic (LAF) insulator[4,5], isospin polarized metals[4-6], the quantum anomalous Hall effect[7,8], and superconductivity[9] in rhombohedral trilayer ($r$-3LG), tetralayer ($r$-4LG) or pentalayer graphene ($r$-5LG). When a moiré superlattice is introduced by aligning rhombohedral graphene with hBN, the intrinsic flat bands are further flattened by the moiré potential. Consequently, Mott insulating states[10], generalized Wigner crystals[11], and integer and fractional quantum anomalous Hall effects[12,13] are observed. Most of the abovementioned correlated states show complex layer-number dependence, violating the simple expectation that the correlation is stronger in thicker layers due to the $E_k \propto p^N$ dispersion. One example is that correlated insulating states at integer fillings are found to be prominent on the hole doping side in $r$-3LG/hBN moiré superlattices, whereas they appear on the electron doping side in thicker graphene systems[10,14,15]. While LAF states and quantum anomalous Hall effects have been observed in specific layer numbers, how the layer number modulates symmetry breaking and screening remains unresolved, particularly in thicker stacks ($N \geq 4$).

Here, we fabricate dual-gated crystalline rhombohedral graphene devices and systematically investigate their electrical transport properties. At the charge neutrality point, we observe multiple symmetry-broken states, including LAF, semimetal and layer-polarized insulator (LPI) phases driven by vertical displacement field, $D$. The critical displacement field $D_c$ for the LAF-to-semimetal transition remains nearly constant from $r$-4LG to $r$-6LG, whereas the $D_c$ for semimetal-to-LPI transition increases monotonically with layer number $N$. Surprisingly, in $r$-6LG, a surface-state-dominated resistance peak and corresponding LLs are observed at the semimetal phase at zero and finite external magnetic fields, which can be understood in terms of enhanced screening effect associated with the increased density of states in thicker rhombohedral graphene multilayers. At high magnetic fields, the dependence of quantum Hall states on $D$ and carrier density $n$ is fully captured by the spin, valley and filling factors at each surface state. Interlayer backscattering between states with opposite chiralities may be responsible for the observed dissipative quantum Hall states.

## II. RESULTS

### A. Broken-symmetry phases and their unconventional transitions

The number of graphene layers is determined by optical contrast after exfoliation onto $SiO_2$/Si substrate. Domains in rhombohedral stacking are identified by infrared scanning near-field optical microscopy[16] and isolated in-situ by atomic force microscopy cutting[17]. To access the full phase diagram in terms of charge carrier density, $n = (C_t V_t + C_b V_b)/e$, and vertical displacement field, $D = (C_b V_b - C_t V_t)/2$, where $C_t$ and $C_b$ are the top- and bottom- gate capacitances per unit area, $V_t$ and $V_b$ are the corresponding gate voltages, rhombohedral graphene is encapsulated by hexagonal boron nitride (hBN) flakes for fabrication into dual-gated devices, as illustrated in Figure 1a. The final stacking order of graphene is confirmed after hBN encapsulation by the phonon-polariton-assisted near-field optical imaging technique [4]. Figure 1b shows a typical *r*-6LG device (*r*-6LG-D1) in a Hall bar geometry achieved by standard e-beam lithography, reactive ion etching and metal deposition.

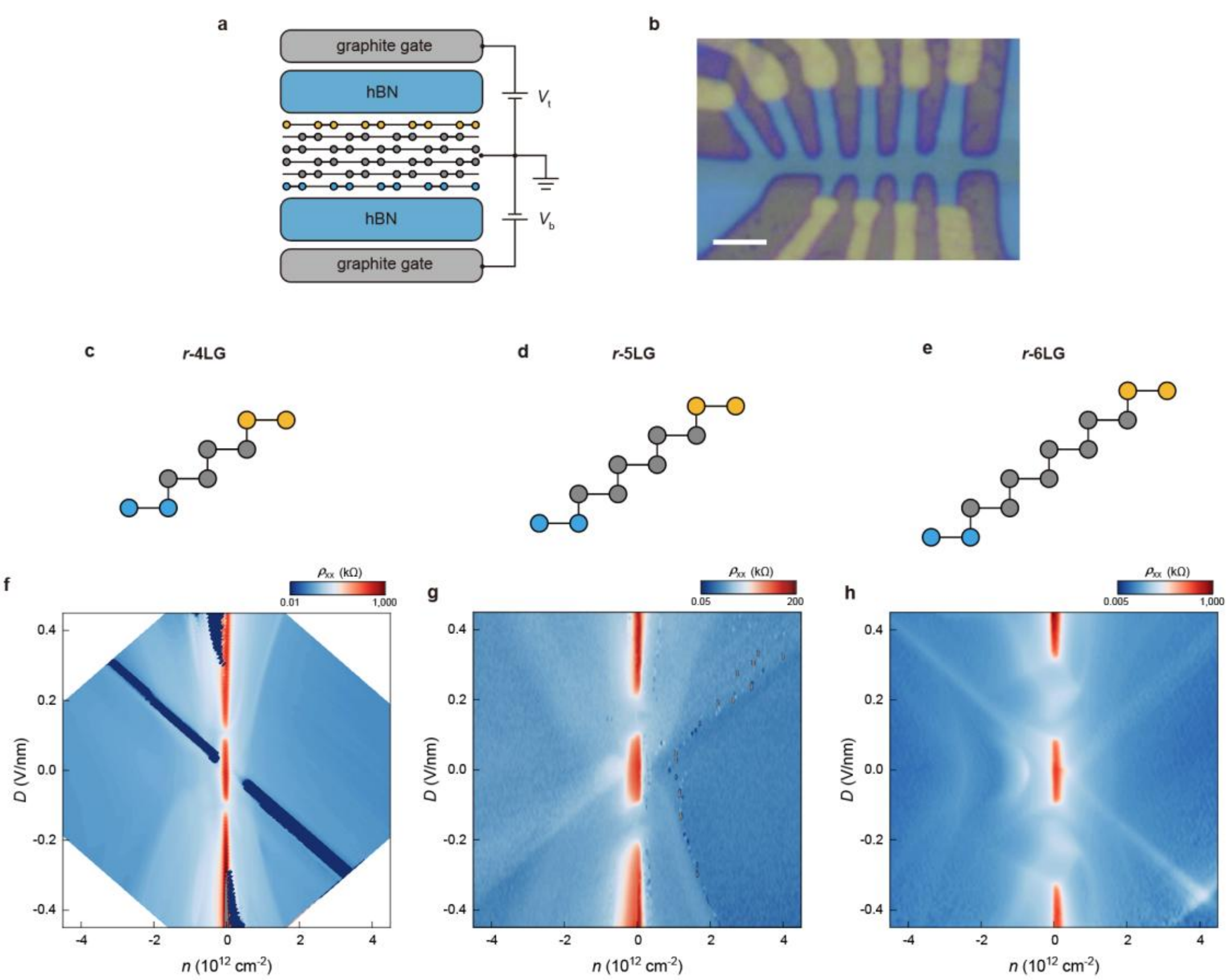

**Figure 1. Rhombohedral 4, 5, 6-layer graphene. a**, Schematic of the dual-gated devices. **b**, Optical image of the *r*-6LG device in Hall bar geometry. Scale bar, 3 µm. **c-e**, Schematic structures of *r*-4LG (**c**), *r*-5LG (**d**), and *r*-6LG (**e**). **f-h**, Longitudinal resistivity maps $\rho_{xx}$ as a function of carrier density *n* and displacement field *D* measured at *T* = 1.5 K for *r*-4LG-D1 (**f**), *r*-5LG-D1 (**g**), and *r*-6LG-D1 (**h**).

Figure 1c-h show illustrations of unit cells and color plots of the measured resistivity $\rho_{xx}$ as functions of *n* and *D* (1.5 K, *B* = 0 T) of *r*-4LG-D1 (c,f), *r*-5LG-D1 (d,g) and *r*-6LG-D1 (e,h), see also Sec. 1 of Supplementary Information for colormaps of *r*-8LG-D1 and other additional devices. From these color plots, three phases with distinct resistivities are resolved at *n* = 0 for different *D* in *r*-4LG through *r*-6LG. The insulating, semimetallic, and insulating phases at different *D* together with transitions between them from *r*-4LG to *r*-6LG are more clearly presented by the line cuts at *n* = 0 measured at different temperatures shown as Fig. 2a (also see Sec. 2 of Supplementary Information for data of *r*-8LG-D1 and additional *r*-5LG devices).

Note that alignment with hBN can dramatically modify the onset $D$ of these phases[14,18]; therefore, only moiréless samples are included in this study (see Fig. S1 for zoomed-out colormaps). Single-particle band-structure calculations indicate that no band gap (overlapping of valence and conduction bands) exists in $r$-4LG through $r$-6LG. Since quantized longitudinal and Hall resistivities are absent at $n = D = 0$, the observed insulating state can be attributed to the LAF state, in which opposite spins reside on the top (yellow) and bottom (blue) layers, emerging from strong Coulomb interactions of the crystalline flat band, as previously reported in $r$-4LG and $r$-5LG.[4,5]. The spontaneous symmetry breaking associated with the LAF state is evidenced by a phase-transition signature at ~ 20 K in temperature-dependent characteristic (black curve in Fig. 2b, also see Sec. 3 of Supplementary Information), in contrast to LPI state observed at large $D$ (> 0.3 V nm$^{-1}$ for $r$-6LG) at $n = 0$ (blue curve in Fig. 2b).

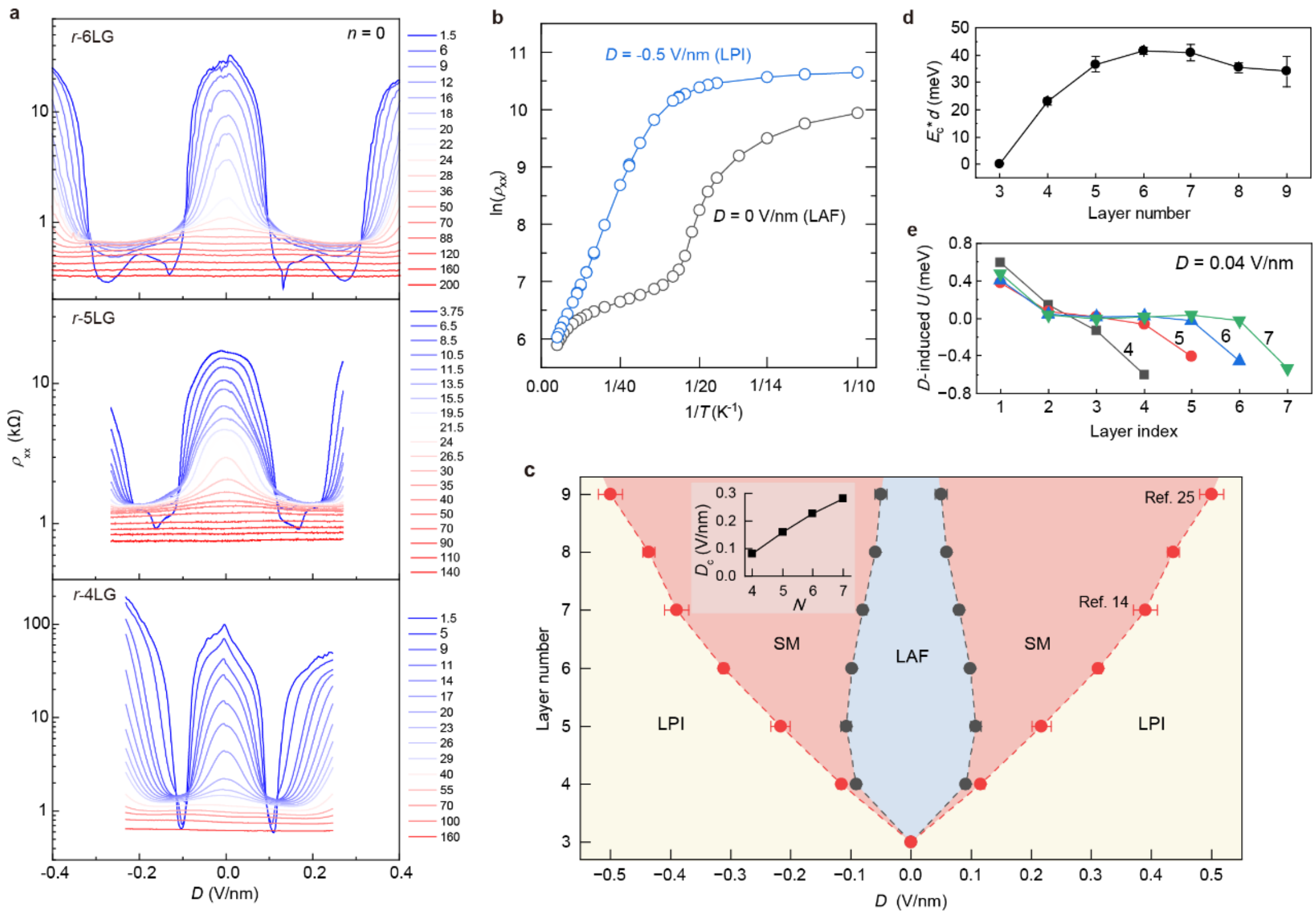

**Figure 2. Layer-dependent phase transitions driven by $D$ at $n$ = 0. a**, $D$-dependent $\rho_{xx}$ of $r$-4LG to $r$-6LG for $n = 0$ at varied temperatures from $T = 1.5$ to 200 K. **b**, $\rho_{xx}$ versus temperature from $T = 1.5$ K to 200 K at $D = 0$ (black) and -0.5 V/nm (blue). **c**, Critical displacement field for phase transitions from layer-polarized insulator (LPI) to semimetal (SM, red circle) and from SM to layer-antiferromagnetic (LAF, black circle) insulator as functions of the layer number in rhombohedral graphene. Inset: $D_c$ of the SM-to-LPI transition for different layer numbers of rhombohedral graphene obtained from a self-consistent tight-binding calculation. **d**, $E_c$*$d$ as a function of layer number. **e**, Calculated layer-resolved screened potential at low displacement field ($D = 0.04$ V/nm in the calculation corresponds to semimetal phase in the experiment).

As $D$ increases, the LAF state is gradually suppressed, followed by a transition to the semimetal state at $D \approx \pm 0.1$ V nm$^{-1}$ for $r$-4LG to $r$-6LG, as shown in Fig. 1f-h, Fig. 2a, and Sec.

2 of Supplementary Information. The semimetal phase is further analyzed and confirmed through fitting of magnetoresistance and Hall resistance for each thickness (see Sec. 4, Supplementary Information).

At larger $D$, the potential difference $\Delta U$ between the top and bottom layers becomes sufficiently large to open a band gap whose size increases linearly with $D$ (Fig. S5), driving the system into the LPI phase (carriers are pushed into the same side) with broken inversion symmetry. Similar behavior is also observed in AB-stacked bilayer graphene[19-22] and $r$-3LG[23,24].

These three distinct phases at charge neutrality are summarized in Fig. 2c, where the phase boundaries are identified from the crossing points (temperature-independent resistivity points) in Fig. 2a. The LAF state (blue region in Fig. 2c) starts to appear from $r$-4LG, maintains an almost constant $D_c$ of approximately $\pm 0.1$ V/nm up to $r$-6LG, narrows from $r$-7LG, and finally disappears in rhombohedral graphite thicker than 4 nm[25]. Based on $D_c$, the interaction strength can be evaluated via the critical potential difference between the top and bottom surface states, $E_c \cdot d$, induced by the out-of-plane electric field at which the correlated LAF gap closes. In this sense, this approach is analogous to the results of transport spectroscopy[20,26] and scanning tunneling spectroscopy[27]. In contrast, thermal activation energy in this case is no longer suitable to assess the correlation strength considering the spontaneous symmetry breaking for the LAF phase (See Sec. 3 of Supplementary Information). As shown in Fig. 2d, the $E_c * d$ increases with layer number and peaks at 6 and 7, indicating maximal interaction strength. At first glance, this behavior is counterintuitive, as stronger Coulomb interactions—and hence a more robust LAF state—would be expected in thicker rhombohedral graphene layers due to the dispersion relation $E_k \propto \pm p^N$. The nonmonotonic layer number dependence of LAF reflects the competition between the density of surface states and screening effects. As the layer number increases, the increased density of surface states enhances interactions; while the enhanced screening in thicker multilayers suppresses correlations[3,28]. This layer-dependent transport behavior is consistent with the recent STM/STS result[27] and previous DFT calculations[29].

At the same time, the onset $D$ of the LPI state increases monotonically with layer number, as shown in Fig. 2c. Within the simplest unscreened model consisting of two flat surface bands with zero band overlap, the size of LPI gap, or $\Delta U$, should be proportional to $D \cdot d$, where $d$ is the thickness of graphene multilayer, regardless of LAF and semimetal phases. Under this assumption, a larger bandgap would be expected in thicker graphene for a given $D$, in contradiction with our experimental observations. A more realistic description must therefore incorporate both the finite band overlap in the semimetal phase and the enhanced screening effects present in thicker rhombohedral graphene multilayers[3]. In this semimetal phase (red regions in Fig. 2c) at finite $D$, the overlap between conduction and valence flat bands at low energy gives rise to substantial carrier populations on the top and bottom surfaces with oppositely charged carriers, leading to strong screening effects in rhombohedral graphene.

To examine the evolution of this screening effect, we performed self-consistent electrostatic tight-binding calculations for rhombohedral multilayer graphene under an applied displacement field[3,30]. The calculated layer-resolved potential profiles are shown in Fig. 2e (also see Sec. 6 of the Supplementary Information). For thicker multilayers, the electrostatic potential remains nearly constant throughout the interior layers, while most of the residual potential drop is concentrated near the outer layers. This behavior indicates that the external displacement field is increasingly screened as the layer number increases.

The enhanced screening has a direct consequence for the stability of the LPI phase. Because only a fraction of the applied displacement field penetrates the multilayer interior, a larger external field is required to generate the layer-polarization potential necessary to drive the transition into the LPI state. Consistent with this picture, the calculated critical displacement field increases monotonically with layer number (inset of Fig. 2c), in good agreement with the experimental trend shown in Fig. 2c.

Quantitative analysis presented in Sec. 6 of the Supplementary Information shows that the screening strength increases systematically with layer number, exceeding 0.94 for $N \geq 6$, and closely tracks the growth of the low-energy spectral weight associated with the surface-derived bands. The consequences of this strong screening can be further examined through the surface-state-dominated transport behavior in the semimetal phase of *r*-6LG.

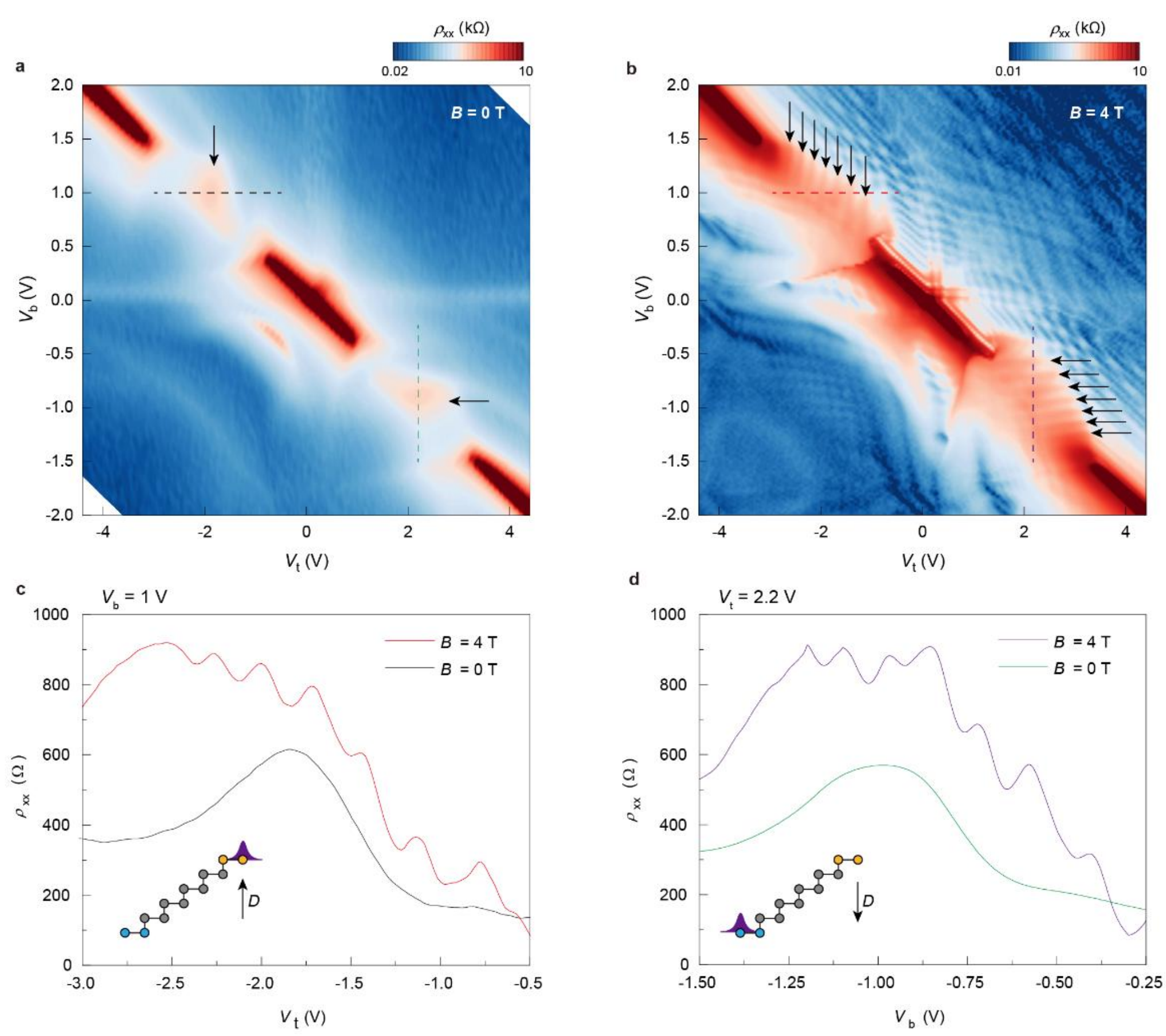

**Figure 3. Surface-state-dominated transport characteristics in *r*-6LG. a**, **b**, longitudinal resistivity $\rho_{xx}$ as a function of top and bottom gate voltages $V_t$ and $V_b$ measured at $B$ = 0 T (**a**) and $B$ = 4 T (**b**), respectively. **c, d,** line-cuts taken in panels **a**, **b**, showing $\rho_{xx}$ as a function of $V_t$ ($V_b$) at negative (positive) external displacement fields. Inset, schematics of the wavefunctions responsible for the change in $\rho_{xx}$ shown in panels **c** and **d**.

## B. Surface-state-dominated transport in *r*-6LG

An additional resistive peak appears in the semimetal phase of *r*-6LG on both positive- and negative-*D* sides (blue curve in Fig. 2a), which is absent in *r*-4LG and *r*-5LG. The peak can be better resolved in the resistivity colormap as a function of $V_t$ and $V_b$ shown in Fig. 3a, where it is marked by black arrows. A more detailed discussion on this resistive peak can be found in Sec. 7 of the Supplementary Information. Notably, the peaks at negative and positive *D* are more sensitive to top and bottom gates, respectively. This single-gate controlled behavior becomes even more evident in a magnetic field when the LLs are developed from the additional peaks. As marked by black arrows in Fig. 3b, the LLs in the positive and negative *D* are developed along $V_t$ and $V_b$ axes, respectively. This dependence on the adjacent gate is directly shown in Fig. 3c and d, which show line cuts along the adjacent gate voltages at zero and finite magnetic fields. These results provide evidence for the surface state exhibiting strong screening of the external electric field. In comparison, the LLs in the other regions depend primarily on the total carrier density despite certain regions also exhibit *D*-dependence due to LL crossings or phase boundaries associated with isospin polarized metals (see Sec. 8 of Supplementary Information).

In contrast to *r*-6LG, surface-state-dominated electron transport is absent in *r*-4LG and *r*-5LG. As discussed in Sec. 9 of the Supplementary Information, over a wide range of perpendicular magnetic fields, the longitudinal resistances of *r*-4LG and *r*-5LG are tunable by both top and bottom gate voltages, and the LLs depend mainly on the total carrier densities. This contrasting behavior between *r*-6LG and thinner graphene is possibly due to the larger density of states of the semimetal phase in *r*-6LG that can contribute stronger screening. The observation of such features requires sufficiently clean devices, but the systematic contrast between *r*-6LG and thinner devices is primarily attributed to the layer-number-dependent screening strength rather than device mobility alone.

## C. Layer-valley locked quantum Hall states

As the magnetic field increases, a signature of a crossover from the single-gate-controlled LL features to dual-gate-controlled quantum Hall states can be observed (see Fig. S12). The surface states in *r*-6LG become more prominent in the form of LLs at high magnetic fields. In the lowest LL of rhombohedral graphene, the valley degree of freedom is locked to the layer index, such that the *K* is locked to one layer (bottom layer), and *K'* to the opposite layer (top layer)[31]. For *r*-6LG, the total filling factor of the lowest LL is ±12[32,33]. Figure 4a shows the colormap of the measured resistance as a function of *D* and the total filling factor $\nu_{tot} = n_{tot}(h/e)/B$, where $h$ is Planck's constant, $e$ is elementary charge (also see colormap of $R_{xy}$ in Fig. S13b). The 24-fold degeneracy of the lowest LL is fully lifted, with multiple LL crossings arising from the interplay of magnetic field, displacement field and Coulomb interactions.

A qualitative understanding of these complex LL crossings can be gained from a simplified scheme of the LL energies as a function of electric field at nonzero magnetic field[14,21,34-36]. In Fig. 4c, each line corresponds to a LL, and T and B represent the top and bottom layers, which are therefore locked to the *K* and *K'* valleys, respectively[31]. At $D = 0$, the spin degeneracy is lifted by the magnetic field. As $D$ increases, the energy of LLs of top (bottom) layer increases (decreases), as illustrated in Fig. 4c. At high magnetic fields, the degeneracy of each LL is lifted, resulting in multiple LL crossings that experimentally appear as the crossing points of the white lines (LLs) in Fig. 4a. We note that the analysis above is also applicable to *r*-7LG and *r*-8LG which provides a reliable method for determining the number of layers of rhombohedral graphene multilayers (see Sec. 12 of Supplementary Information).

Following the guideline of Fig. 4c, we construct the colormap shown in Fig. 4b to label the flavor combinations of the LLs developed on both surfaces, providing a direct counterpart to each experimentally observed state in Fig. 4a[21,35,37]. The states are labelled by four flavors, filling factor, spin, valley, and layer. For instance, (-1*K'*↓, 1*K*↑) represents a total filling factor of 0 state composed of two LLs residing on top and bottom layers: a top-layer LL with $\nu = -1$, *K'* valley, and spin down, and a bottom-layer LL with $\nu = 1$, *K* valley, and spin up.

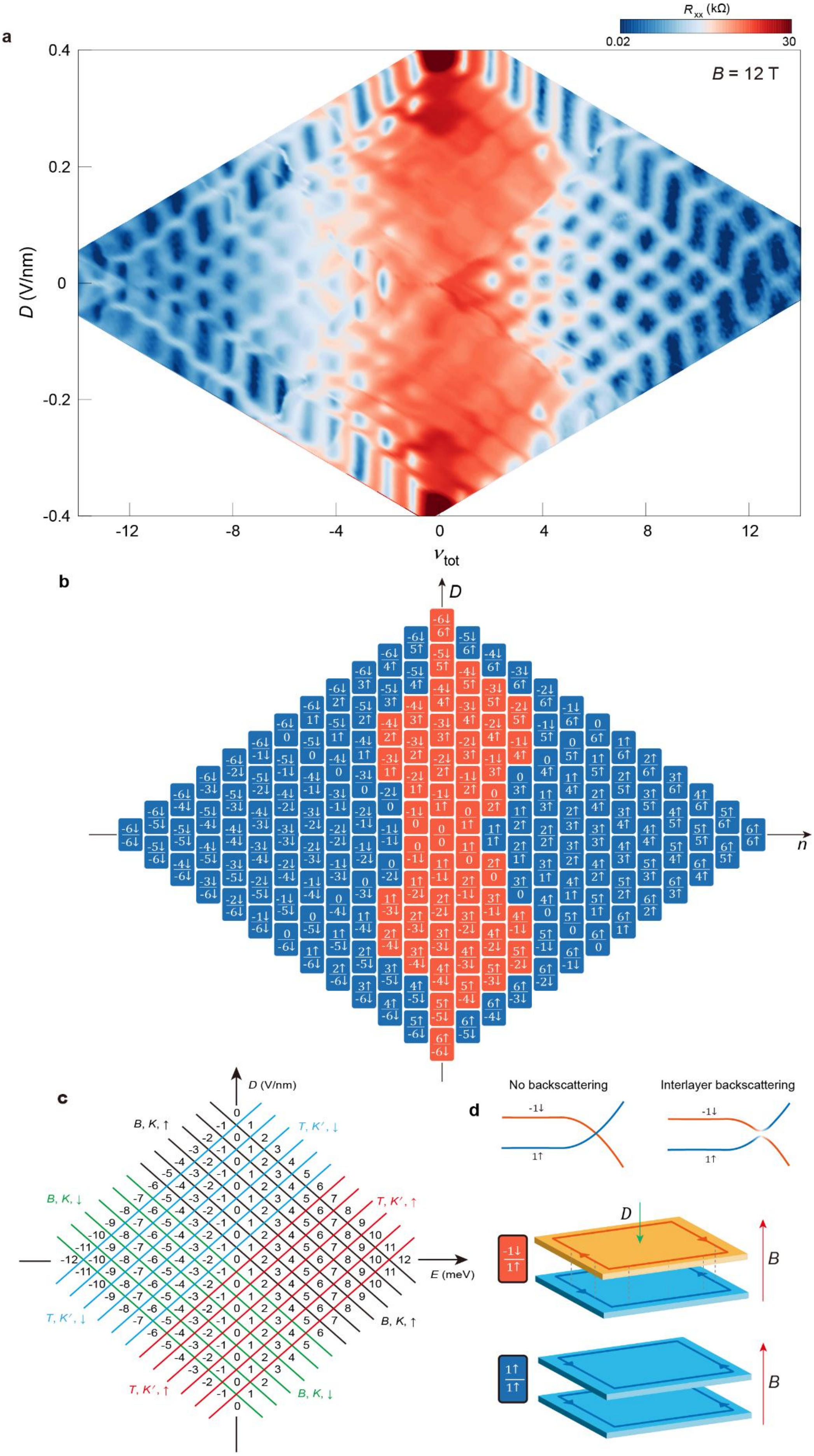

**Figure 4. Landau levels in *r*-6LG. a**, longitudinal resistivity $\rho_{xx}$ as a function of displacement field $D$ and total filling factor $\nu$ measured at $B$ = 12 T. **b**, Schematic diagram of the evolution of the zeroth LLs with displacement field. In the diagram, each square represents a state with specific filling factor, spin and valley. **c**, Schematic illustration of LL crossings driven by displacement field under a constant magnetic field. 'T': top layer; 'B': bottom layer. The information of energy difference between LLs is ignored for simplicity. **d**, Schematic illustration of the edge dispersion of the LLs without (top-left panel) and with (top-right panel) interlayer backscattering. A gap the opened at the edge when interlayer backscattering is present. Cartoons of QH edge states residing on the top and bottom layers, when $\nu_{tot}$ = 0, finite $D$ (middle panel); $\nu_{tot}$ = 2, $D$ = 0 (bottom panel).

We note that in Fig. 4a, there are two diamond-shaped regions showing large $R_{xx}$ even when the total filling factor lies within an LL gap. By comparison with Fig. 4b, we find that these dissipative quantum Hall states correspond to LLs on the top and bottom surfaces with opposite signs of filling factors (counter-propagating edge modes) and opposite spins. In contrast, $R_{xx}$ decreases to nearly zero when the LLs on top and bottom surfaces possess the same chirality and spin (see Sec. 11 of Supplementary Information for a more detailed analysis). These contrasting behaviors can be explained by interlayer backscattering between counter-propagating edge states on the top and bottom surfaces (Fig. 4d). The hole-like and electron-like LLs cross and open a gap in the presence of interlayer backscattering, leading to the vanishing of the quantum Hall state. We note that this suppression of quantum Hall effect occurs only at low total filling factors (no greater than 4) and in the presence of finite $D$, suggesting that interlayer backscattering is sensitive to Coulomb interactions.

## III. DISCUSSION

Our results reveal several notable layer-dependent effects:

1. LAF stability vs. layer thickness: While Coulomb interactions strengthen with $N$, the characteristic displacement-field scale associated with the LAF phase reaches a maximum at $N$ = 6 and 7 and decreases in thicker layers. This behavior suggests that band reconstruction (e.g., trigonal warping) outweighs interaction effects beyond a critical thickness.

2. Screening-band-overlapping-engineered LPI: The increase in $D_c$ for the LPI transitions with $N$ underscores the combined role of finite band overlap and enhanced electronic screening, with the latter increasingly associated with surface-derived low energy states in thicker multilayers.

3. Surface states in the semimetal phase: Surface states tuned predominantly by the adjacent gate voltage are observed, which are likely responsible to the enhanced screening in the semimetal phase of *r*-6LG.

In addition, the gate-tunable surface states in *r*-6LG provide a platform for designer quantum Hall bilayers, where interlayer backscattering can be electrically switched. Unlike twist-engineered systems, our approach leverages intrinsic layer number—a scalable parameter suitable for wafer-scale device integration.

This study of the evolution of correlated phases and their transitions with layer thickness provides important insight into the correlation strength in rhombohedral graphene multilayers for the search of versatile correlated and topological phases. The surface states arising from screening effects together with their LLs host rich degeneracies and transitions tunable through multiple parameters including Coulomb interactions, displacement field and carrier density. These systems therefore offer an attractive platform for studying quantum Hall double layers and novel fractional quantum Hall states.

## APPENDIX: METHODS

### 1. Sample fabrication

Graphene and hBN flakes are mechanically exfoliated onto $SiO_2$ (285 nm)/Si substrates, and the layer numbers are identified by optical contrast and atomic force microscope (AFM) topographic characterizations. The rhombohedral graphene domain is identified using scanning near-field infrared microscope and isolated in-situ using the AFM cutting. The van der Waals heterostructure is fabricated by a standard polycarbonate-based dry transfer process. First, a bottom gate structure is fabricated by putting down a hBN flake on a graphite flake. Then, the heterostructure is annealed at 300 ºC in hydrogen and argon. Next, the top graphite, top hBN and rhombohedral graphene are picked up successively and dropped on the prepared bottom gate. After the wash-off process with chloroform, standard e-beam lithography is used to define Hall bar contacts, and Cr/Au electrodes are deposited by e-beam evaporation.

### 2. Transport measurements

The devices are measured in a 1.5 K-base-temperature Oxford variable-temperature insert system. Stanford Research Systems SR830/860 and Guangzhou Sine Scientific Instrument OE1201 lock-in amplifiers with an alternating current of 5 nA-10 nA at a frequency of 17.77 Hz in combination with a 100 MΩ resistor are used to measure the resistivity. Keithley 2400 and 2401 source meters are used to apply the gate voltages.

### 3. Self-consistent electrostatic tight-binding calculations

We evaluate the self-consistent electrostatic tight-binding model in a low-energy continuum form expanded around the $K$ valley. In the layer-sublattice basis $\Psi_{\mathbf{k}} = \left(\psi_{A_1,\mathbf{k}}, \psi_{B_1,\mathbf{k}}, \psi_{A_2,\mathbf{k}}, \psi_{B_2,\mathbf{k}}, \dots, \psi_{A_N,\mathbf{k}}, \psi_{B_N,\mathbf{k}}\right)^T$, the single-particle Hamiltonian is written as

$$\hat{H} = \sum_{\mathbf{k}} \Psi_{\mathbf{k}}^{\dagger} \mathcal{H}_K^{(N)}(\mathbf{k}) \Psi_{\mathbf{k}}. \qquad (1)$$

Here $\mathbf{k} = (k_x, k_y)$ is measured from the $K$ point, and the continuum Hamiltonian is

$$\mathcal{H}_K^{(N)}(\mathbf{k}) = \begin{pmatrix} h(\mathbf{k})+D_1 & T_1(\mathbf{k}) & T_2 & 0 & \dots & 0 \\ T_1^\dagger(\mathbf{k}) & h(\mathbf{k})+D_2 & T_1(\mathbf{k}) & T_2 & \ddots & \vdots \\ T_2^\dagger & T_1^\dagger(\mathbf{k}) & h(\mathbf{k})+D_3 & T_1(\mathbf{k}) & \ddots & 0 \\ 0 & T_2^\dagger & T_1^\dagger(\mathbf{k}) & h(\mathbf{k})+D_4 & \ddots & T_2 \\ \vdots & \ddots & \ddots & \ddots & \ddots & T_1(\mathbf{k}) \\ 0 & \dots & 0 & T_2^\dagger & T_1^\dagger(\mathbf{k}) & h(\mathbf{k})+D_N \end{pmatrix}, \quad (2)$$

with

$$h(\mathbf{k}) = \begin{pmatrix} 0 & v\pi^\dagger \\ v\pi & 0 \end{pmatrix}, \qquad \pi \equiv k_x + ik_y, \quad (3)$$

$$T_1(\mathbf{k}) = \begin{pmatrix} v_4\pi^\dagger & v_3\pi \\ t_1 & v_4\pi^\dagger \end{pmatrix}, \qquad T_2 = \frac{t_2}{2}\begin{pmatrix} 0 & 1 \\ 0 & 0 \end{pmatrix}, \qquad D_l = \begin{pmatrix} U_l + \Delta'\delta_{l1} & 0 \\ 0 & U_l + \Delta'\delta_{lN} \end{pmatrix}. \quad (4)$$

The velocity parameters are defined by $v_i = (\sqrt{3}a/2\hbar)t_i$ for $i = 0,3,4$, where $a$ is the graphene lattice constant. The hopping amplitudes follow the standard Slonczewski-Weiss-McClure convention for multilayer graphene[38]. The parameters used in the present calculations are $t_0 = 3.1$ eV, $t_1 = 0.38$ eV, $t_2 = -0.015$ eV, $t_3 = -0.29$ eV, $t_4 = -0.141$ eV, and $\Delta' = 0.0105$ eV. The total carrier density is fixed to $n_{\text{tot}} = 0$, while $N = 4,5,6,7$ and the displacement field $D$ are varied to obtain the gap evolution, screened layer-potential profiles, and layer-resolved observables shown in the main text.

The electrostatic potentials are determined self-consistently together with the band occupation. The potential drop between neighboring layers satisfies the discrete Gauss-law relation[39,40]

$$U_{l+1} - U_l = -\frac{e^2 d}{\varepsilon_\perp \varepsilon_0}\left(n_b + \sum_{j\le l} n_j\right), \quad (5)$$

where $d = 3.35$ Å is the interlayer spacing, $\varepsilon_\perp = 4$ is the out-of-plane dielectric constant, $n_j$ is the charge density on layer $j$, and $n_t$ and $n_b$ are the top- and bottom-gate charge densities. These quantities satisfy

$$D = \frac{e(n_b - n_t)}{2\varepsilon_0}, \qquad n_{\text{tot}} = \sum_l n_l = -(n_t + n_b). \quad (6)$$

Here $e$ is the elementary charge and $\varepsilon_0$ is the vacuum permittivity. At fixed $(n_{\text{tot}}, D)$, the Hamiltonian is diagonalized to obtain the band energies $E_{n\mathbf{k}}$ and eigenvectors, from which the layer charge densities are evaluated and the potentials $U_l$ are updated iteratively until self-consistency is reached. The screened displacement-field-induced potential shown in the manuscript is defined as the change in the self-consistent layer potential relative to its zero-displacement-field value.

The layer-resolved density of states is defined as

$$\nu_l(\mu) = \sum_{n,\mathbf{k}} w_{n\mathbf{k}}^{(l)}\,\delta(E_{n\mathbf{k}} - \mu), \quad (7)$$

where $n$ labels the band index, $\mu$ is the chemical potential, and $w_{n\mathbf{k}}^{(l)}$ is the projection weight of the eigenstate $(n, \mathbf{k})$ onto layer $l$.

**ACKNOWLEDGMENTS**

The work is supported by Quantum Science and Technology-National Science and Technology Major Project (grant no. 2025ZD0300500) and NSF of China (grant no. 12350005). G.C. acknowledges support from National Key Research Program of China (grant no. 2021YFA1400100), NSF of China (grant no. 12550403), Shanghai Science and Technology Innovation Action Plan (grant no. 24LZ1401100), Scientific Research Innovation Capability Support Project for Young Faculty (no. ZY2025064) and Shuguang Program supported by Shanghai Education Development Foundation and Shanghai Municipal Education Commission. B.L. acknowledges support from the Development Scholarship for Outstanding PhD of Shanghai Jiao Tong University. K.L. acknowledges support from the National Natural Science Foundation of China (Grant No.124B2071). K.W. and T.T. acknowledge support from the JSPS KAKENHI (Grant Numbers 20H00354, 21H05233 and 23H02052) and World Premier International Research Center Initiative (WPI), MEXT, Japan. Part of the measurement was performed in Oxford Instrument Nanoscience Shanghai Demo Laboratory.

**AUTHOR CONTRIBUTIONS**

G.C. conceived and supervised the project. B.L., J.Z. and K.L. fabricated the devices and performed the transport measurements with the assistance of Y. S., S.L. and S.W.. K.W. and T.T. grew hBN bulk crystals. B.L. and J.Z. performed the near-field scanning infrared microscope and AFM measurements. Y.R., and W.L. conducted theoretical calculations. B.L., J.Z. and G.C. analyzed the data. B.L. and G.C. wrote the paper with input from all authors.

**DATA AVAILABILITY**

The data supporting the findings of this study are available within this paper and its Supplementary Information files or from the corresponding authors upon request.

Supplementary Information for

# Layer-Number-Controlled Symmetry Breaking and Surface-State Transport in Rhombohedral Graphene Multilayers

Bosai Lyu[1,2†], Jian Zheng[1†], Kai Liu[1†], Yulu Ren[1], Size Wu[1], Yating Sha[1], Shuhan Liu[1], Youngju Park[3], Kenji Watanabe[4], Takashi Taniguchi[5], Jinfeng Jia[1], Zhiwen Shi[1], Jeil Jung[3,6], Weidong Luo[1], Guorui Chen[1*]

[1] State Key Laboratory of Micro-nano Engineering Science, Key Laboratory of Artificial Structures and Quantum Control (Ministry of Education), Tsung-Dao Lee Institute and School of Physics and Astronomy, Shanghai Jiao Tong University, Shanghai, People's Republic of China.

[2] Department of Quantum Matter Physics, University of Geneva, 24 Quai Ernest Ansermet, CH 1211 Geneva, Switzerland.

[3] Department of Physics, University of Seoul, Seoul, Korea.

[4] Research Center for Electronic and Optical Materials, National Institute for Materials Science, 1-1 Namiki, Tsukuba 305-0044, Japan.

[5] Research Center for Materials Nanoarchitectonics, National Institute for Materials Science, 1-1 Namiki, Tsukuba 305-0044, Japan.

[6] Department of Smart Cities, University of Seoul, Seoul, Korea.

[†]These authors contributed equally to this work.

[*]Contact author: chenguorui@sjtu.edu.cn

## Content

**11. Dissipative and non-dissipative quantum Hall states at $B$ = 12T**

**12. LL crossing patterns of $r$-6LG and $r$-8LG at high magnetic fields**

## 1. Additional colormaps of *r*-5LG, *r*-6LG, and *r*-8LG devices at $B$ = 0T.

In addition to devices shown in Figure 1, more devices are fabricated and characterized (Fig. S1). For the example of *r*-5LG (Fig. S1(a), S1(b), and Fig. 1(g)), all the devices have shown consistent critical *D*s, indicating the robust phase transition behaviors reported in this study. Noted that alignment with hBN can significantly modify the onsets *D* of the LAF and LPI phases[1]. To rule out the moiré-related effect, zoom-out maps (see Fig. S1(e-f), till $n \sim 10^{13}$ cm$^{-2}$) are routinely taken for confirmation the absence of moiré-related resistance peaks.

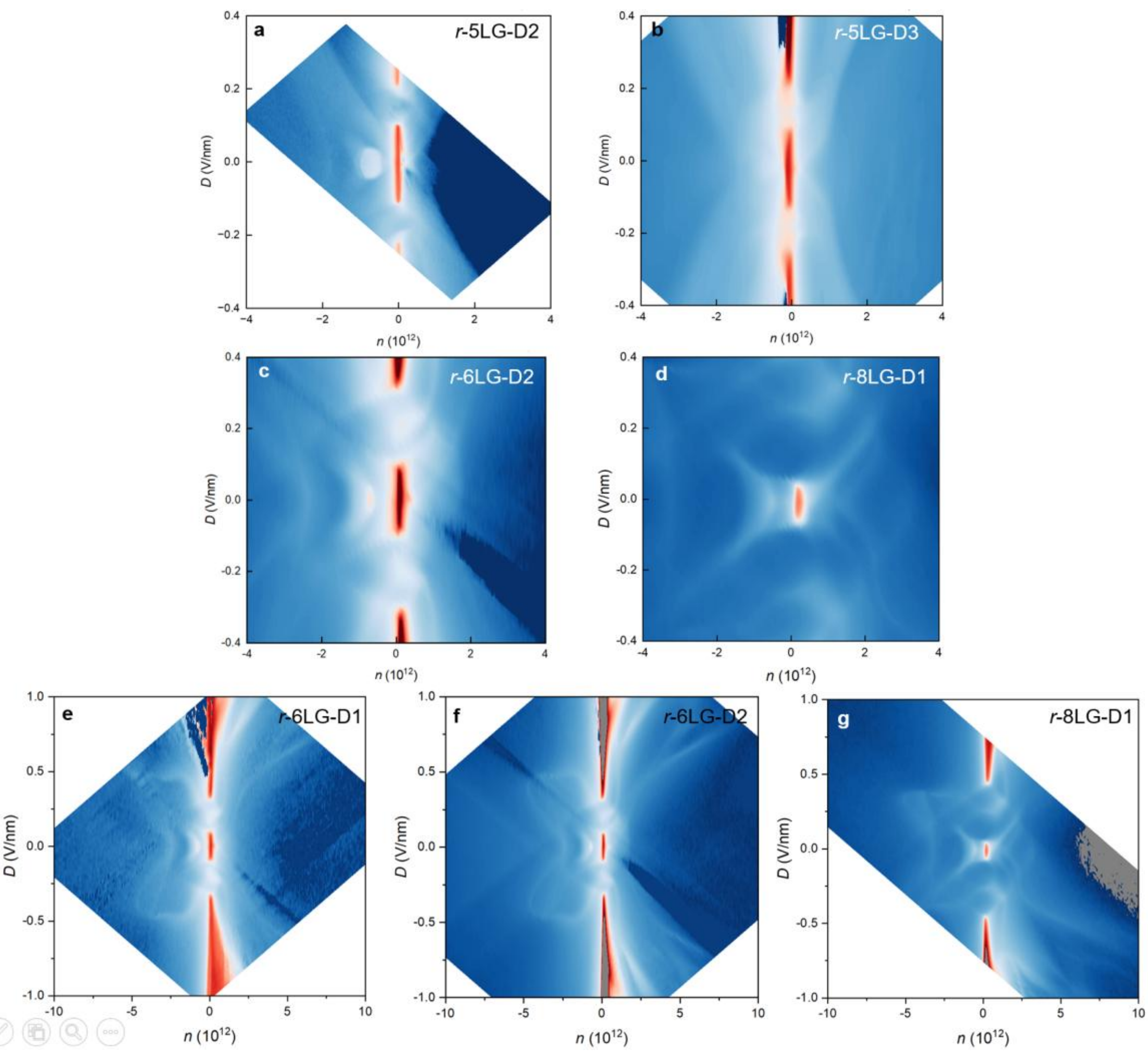

Figure S1. Additional colormaps of *r*-5LG (a and b), *r*-6LG (c and e), and *r*-8LG (d and g) devices at $B$ = 0T. Zoomed-in maps (a-d); Large-range maps (e-g).

## 2. *D*-dependent $\rho_{xx}$ of two additional *r*-5LG and a *r*-8LG for *n* = 0 at varied temperatures.

In addition to the data of pentalayer sample (*r*-5LG-D2) shown in Fig. 2(a), the other two pentalayer devices also exhibit critical *D* values (Fig. S2), that agree well with existing data, with small enough variations. They are also consistent with previous results on moiréless pentalayer graphene[2]. In the same *D* range, LPI states are not observed in *r*-8LG due to its much higher onset *D*, which is attributed to increasing band overlap of the semimetal phase and the enhanced screening effects present in thicker rhombohedral graphene multilayers.

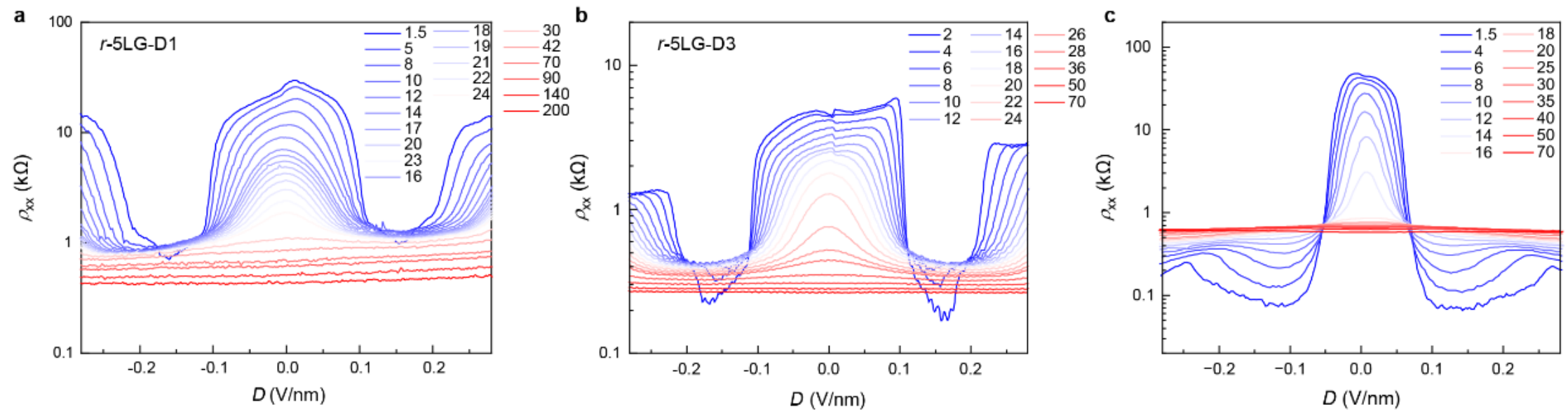

Figure S2. Temperature-dependent data at *n* = 0 of two additional *r*-5LG samples (a and b), and a *r*-8LG (c) sample.

## 3. Temperature dependent behavior of LAF phase for *r*-4-6LG and *r*-8LG.

To understand better the layer-dependence of LAF states, temperature dependent transport behaviors at the charge neutrality $n$ = 0 of *r*-4-6LG and *r*-8LG are discussed in this section. Owing to the correlated nature, the gap size of the LAF states cannot be properly assessed by thermal activation estimations, as discussed in the following.

1, The LAF insulator state is a correlated gap from spontaneous symmetry breaking of graphene in contrast to the single-particle gap, such as LPI. Consequently, resistance of LAF shows complicated temperature dependent behavior in draft contrast to the typical thermally activated behavior observed in LPI at large $D$ (e.g., $D$ = -0.5 V/nm in Fig. 2(b)). As shown in Arrhenius plots in Fig. S3(a), the thermal excitation data of LAF is always nonlinear over the entire temperature range, e.g. there is no linear regime spanning more than one order of magnitude in resistivity, which implies that temperature dependent gap size[2-4].

2, Contradiction between the extracted gap and the onset temperature of the LAF state. In addition to the *r*-4LG to *r*-6LG samples, we fabricated a *r*-8LG device, whose Arrhenius plot is shown in Fig. S3(a). From the Arrhenius plot of the LAF of *r*-8LG, a gap size of ~37 meV can be extracted, which is larger than those of *r*-4LG to *r*-6LG. However, the onset temperature of LAF state—identified by the abrupt resistance upturn in Fig. S3(c) and S3(d), is only ~15 K for *r*-8LG, compared to ~24 K for *r*-4LG to *r*-6LG. Therefore, for the LAF insulating state, the onset temperature and the extracted gap are not consistent.

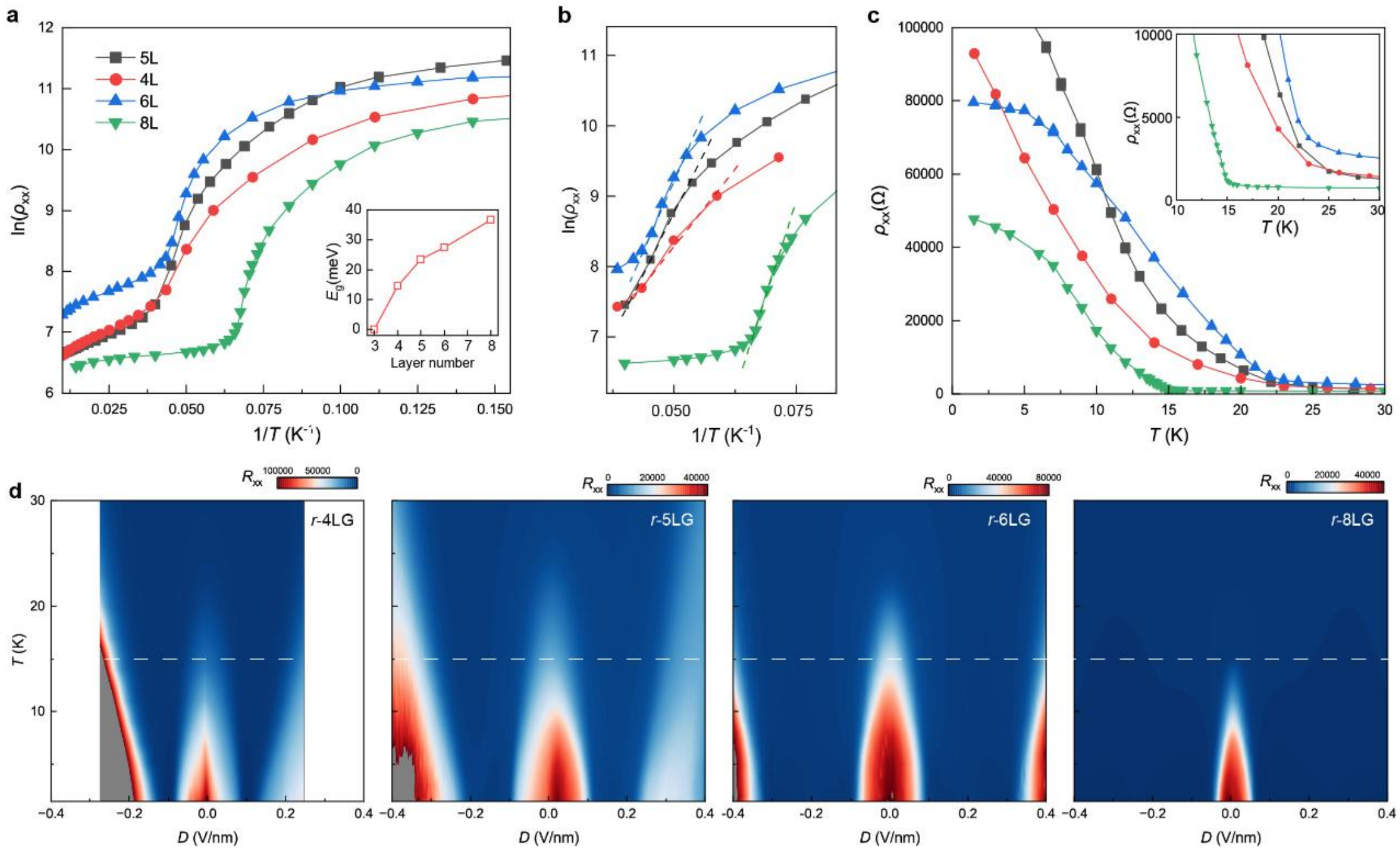

Figure S3. (a) Arrhenius plot: ln($\rho_{xx}$) versus temperature of the *r*-4-6LG and *r*-8LG at $n$ = $D$ = 0. Inset, extracted gap size as a function of number of layers. (b) linear fitting for extraction of thermal excitation gap. (c) $\rho_{xx}$ -$T$ plot at $n$ = $D$ = 0 for *r*-4-6LG and *r*-8LG samples. Inset: zoom-in of the upturn regime. (d) D-T color plots at $n$ = 0 for *r*-4-6LG and *r*-8LG samples.

## 4. Analysis on magneto-transport in the semimetal phase

Positive magnetoresistances are observed in semimetal phases in *r*-4-6LG. For in-depth understanding of layer-dependent magneto-transport in the semimetal phase, we extract electron and hole densities and their mobilities for each thickness via fitting of the longitudinal and Hall resistivity data (Fig. S4) using the following equations based on the two-band model:

$$\rho_{xx} = \frac{1}{e}\frac{n\mu_n + p\mu_p + (n\mu_n + p\mu_p)\mu_n\mu_p B^2}{(n\mu_n + p\mu_p)^2 + (p-n)^2\mu_n^2\mu_p^2B^2}$$

$$\rho_{xy} = \frac{1}{e}\frac{(p\mu_p^2 - n\mu_n^2)B + (p-n)\mu_n\mu_p B^2}{(n\mu_n + p\mu_p)^2 + (p-n)^2\mu_n^2\mu_p^2B^2}$$

Where $n$ and $p$ are the electron and hole densities, $\mu_n$ and $\mu_p$ are electron and hole mobilities, respectively. As shown in Fig. S4 below, the experimental data are well described by the model, yielding the electron and hole densities and mobilities for the semimetal phases of *r*-4LG, *r*-5LG, and *r*-6LG. Comparing to thinner graphene samples, *r*-6LG exhibits carrier densities (~$10^{12}$, comparing to $10^{10}$ in *r*-4LG and $10^{11}$ in *r*-5LG) that are approximately an order of magnitude larger and substantially lower mobilities, resulting in significantly smaller magnetoresistance, arising from an increased band overlap, as well as to its flatter bands. Since rhombohedral multilayer graphene poses nearly flat bands localized at the top and bottom surfaces, with the energy dispersion relation, $E_k \propto p^N$, where $E_k$ is kinetic energy, $p$ is momentum and $N$ is the number of layers. As $N$ increases, the bands become flatter, leading to enhanced interaction effects, consistent with the stronger correlations observed in *r*-6LG.

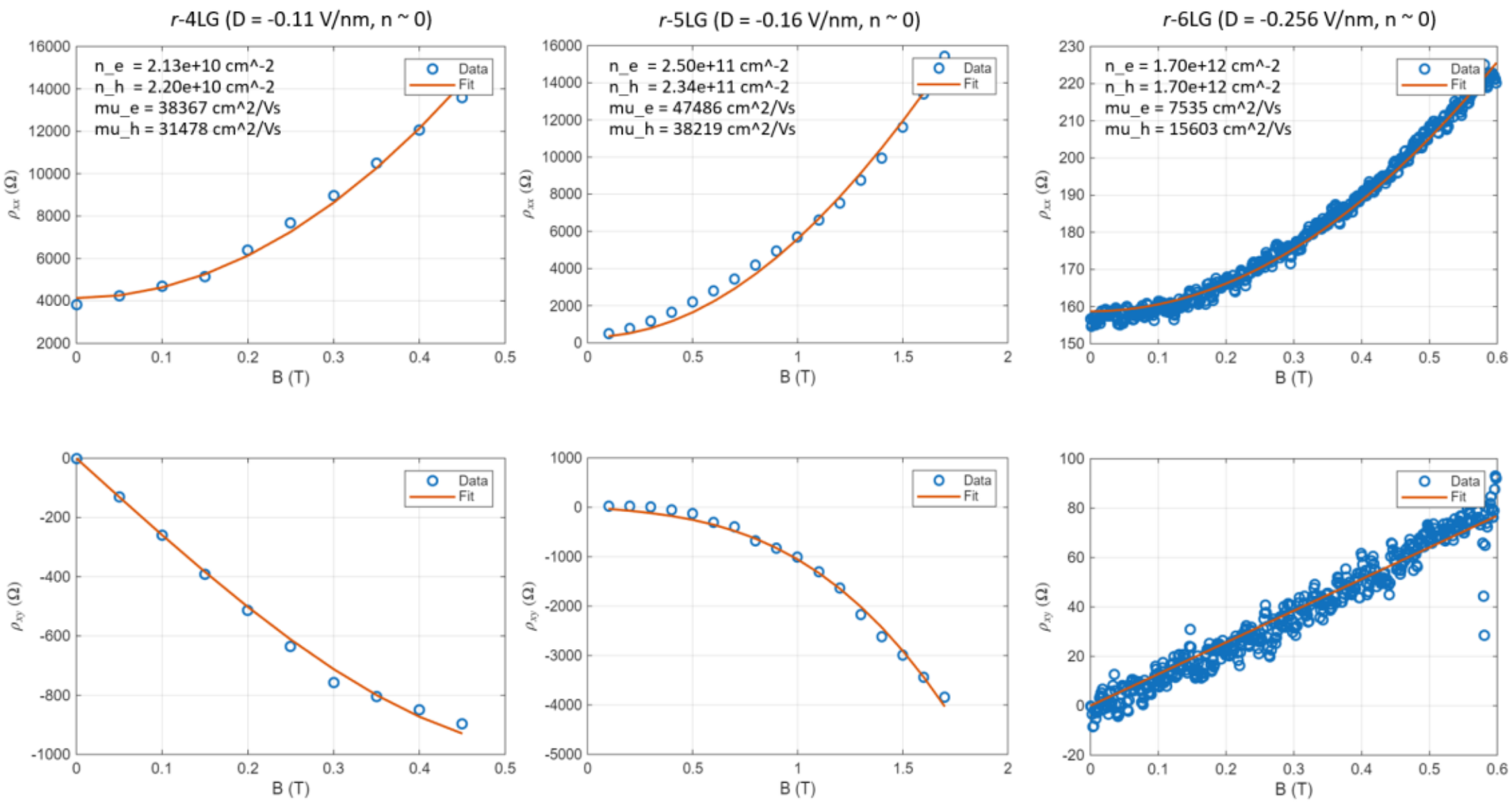

Figure S4. Extraction of carrier densities and mobilities via fitting of magnetoresistance and Hall resistance.

## 5. Layer-number and displacement-field-dependent LPI gap

Gap size Δ of the LPI state can be extracted from the fitting of linear regime of the Arrhenius plot. As shown in Fig. S5, for 4L to 6L, Δ all exhibits linear dependence on *D*, expected for tunable nature of LPI by out-of-plane electric field. However, the layer-number dependent onset *D* of LPI is determined by the interplay of screening effect and band overlap which also evolve with number of layers for rhombohedral graphene multilayers.

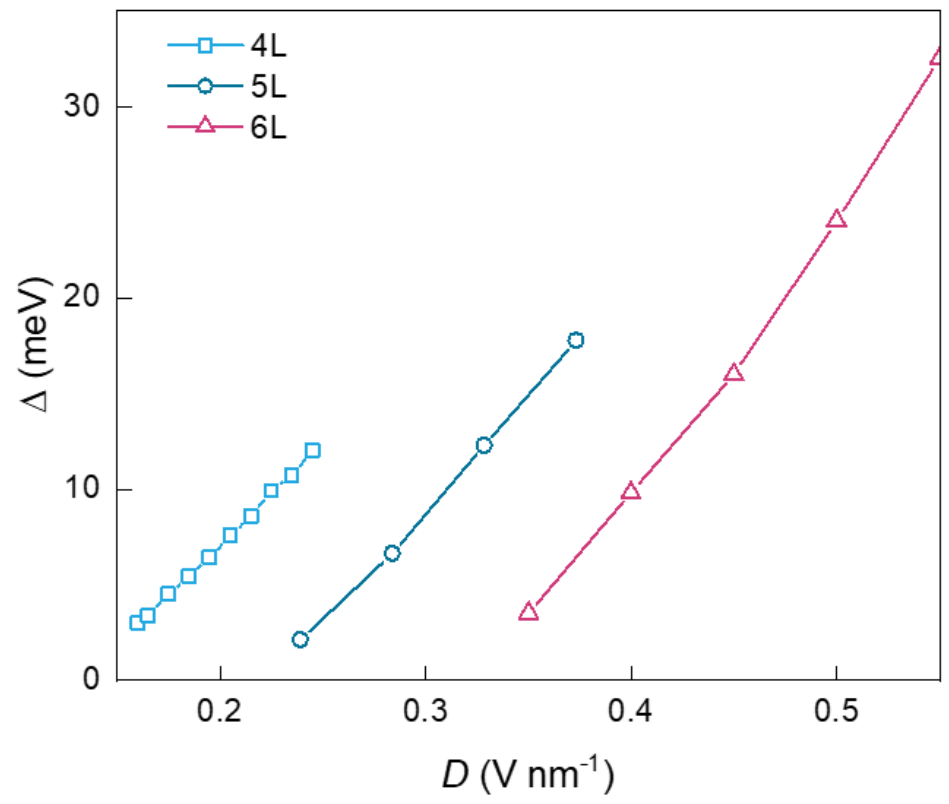

Figure S5. The LPI gap Δ vs *D* for different layers.

## 6. Quantitative Analysis of Electrostatic Screening in Rhombohedral Multilayer Graphene

To further elucidate the origin of the enhanced screening discussed in the main text, we analyze the self-consistent electrostatic response of rhombohedral multilayer graphene under an applied displacement field.

### A. Layer-resolved electrostatic potential

Figures S6 show the layer-resolved electrostatic potentials obtained from the self-consistent calculations for different layer numbers and displacement fields. For a fixed external displacement field, the electrostatic potential becomes progressively flatter in the interior layers as the layer number increases. In thicker multilayers, most of the residual potential drop is confined to the outer layers, indicating that the external field is strongly screened before penetrating into the bulk interior.

This behavior persists over a broad range of displacement fields and demonstrates that the screening response becomes increasingly effective in thicker rhombohedral graphene multilayers.

### B. Screening factor

To quantify the strength of this screening response, we define a screening factor

$$S = 1 - \frac{\Delta U_{SC}}{\Delta U_{unscreened}}$$

where $\Delta U_{SC}$ is the total self-consistently screened potential drop across the multilayer stack, and $\Delta U_{unscreened} = D(N-1)d/\varepsilon_{\perp}$ is the corresponding potential drop expected in the absence of electronic charge redistribution. Here $d$ is the interlayer spacing and $\varepsilon_{\perp} = 4$ is the fixed out-of-plane dielectric constant used in all calculations.

Because $\Delta U_{unscreened}$ already incorporates the background dielectric response, $S$ directly measures the additional screening arising from self-consistent electronic charge redistribution. The same value of $\varepsilon_{\perp}$ is used for all layer numbers, allowing the evolution of $S$ with $N$ to be directly compared.

The calculated screening factor is shown in Fig. S7. For a fixed displacement field, $S$ increases systematically with layer number and exceeds 0.94 for $N \geqslant 6$. In contrast, $S$ decreases gradually with increasing displacement field for a fixed layer number, reflecting the progressive weakening of screening as the system approaches the layer-polarized insulating state.

These results demonstrate that the enhancement of screening with increasing layer number cannot be attributed solely to the increased geometric thickness of the multilayer stack. Instead, it originates from the increasingly effective electronic charge redistribution in thicker rhombohedral multilayers.

### C. Layer-resolved density of states

To identify the microscopic origin of the enhanced screening, we calculate the layer-resolved density of states at zero displacement field. As shown in Fig. S8, the low-energy spectral weight becomes increasingly concentrated on the outermost layers as the layer number increases, particularly between $N = 4$ and $N = 6$.

The evolution of the low-energy spectral weight closely correlates with the increase of the screening factor shown in Fig. S7. Since the electronic compressibility is determined by the low-energy density of states, this result indicates that the enhanced screening is associated with the growing contribution of the low-energy surface-derived bands.

Taken together, the layer-resolved potential profiles, the screening factor analysis, and the density-of-states calculations provide a consistent microscopic picture of the screening mechanism. While the background dielectric response is included through the fixed dielectric constant $\varepsilon_{\perp}$, the enhanced screening observed in thicker rhombohedral multilayers is primarily driven by the increasing electronic compressibility of the low-energy surface-derived states.

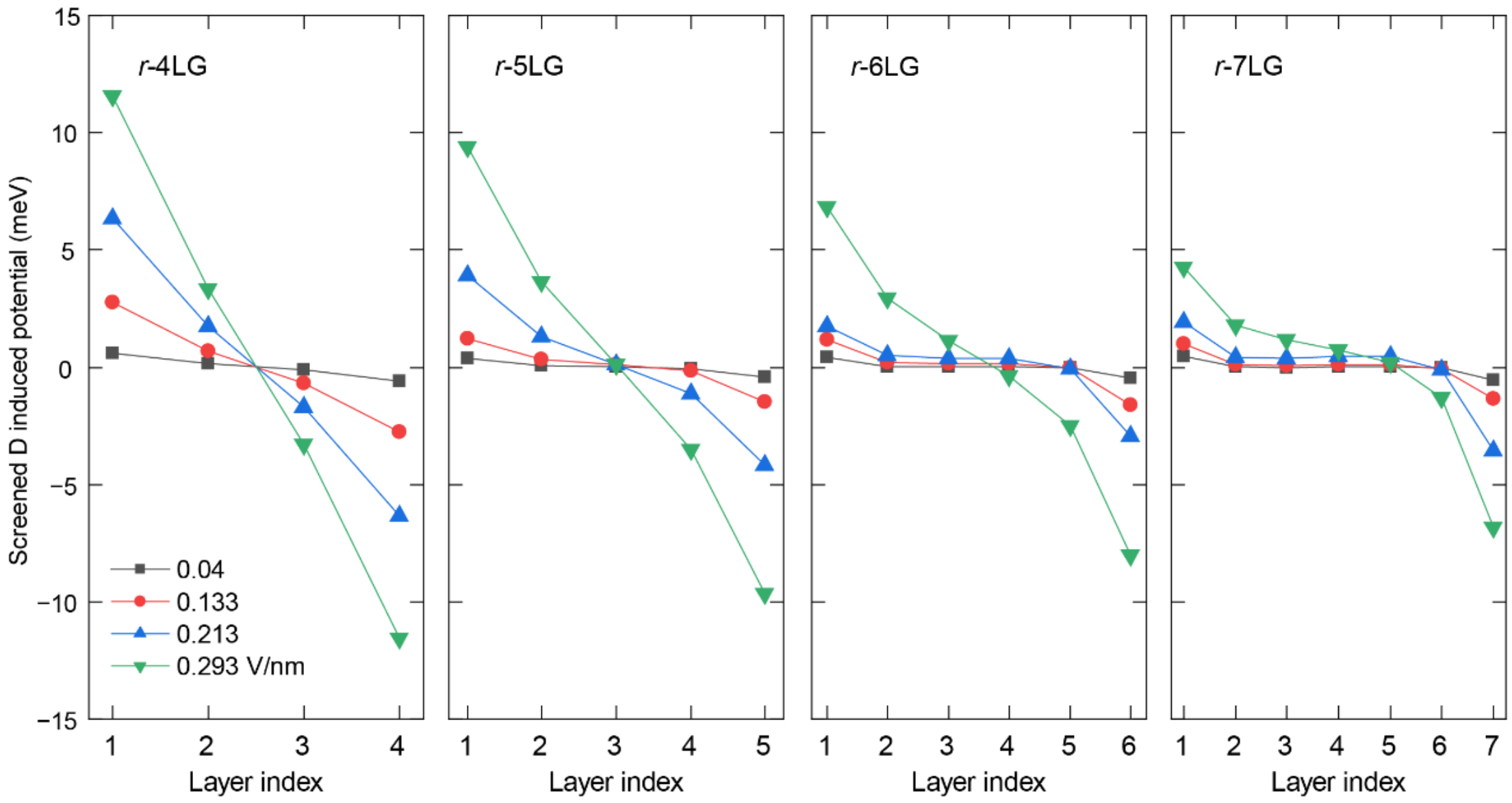

Figure S6. The layer-resolved screened potential in rhombohedral 4-7 layered graphene at different low displacement fields.

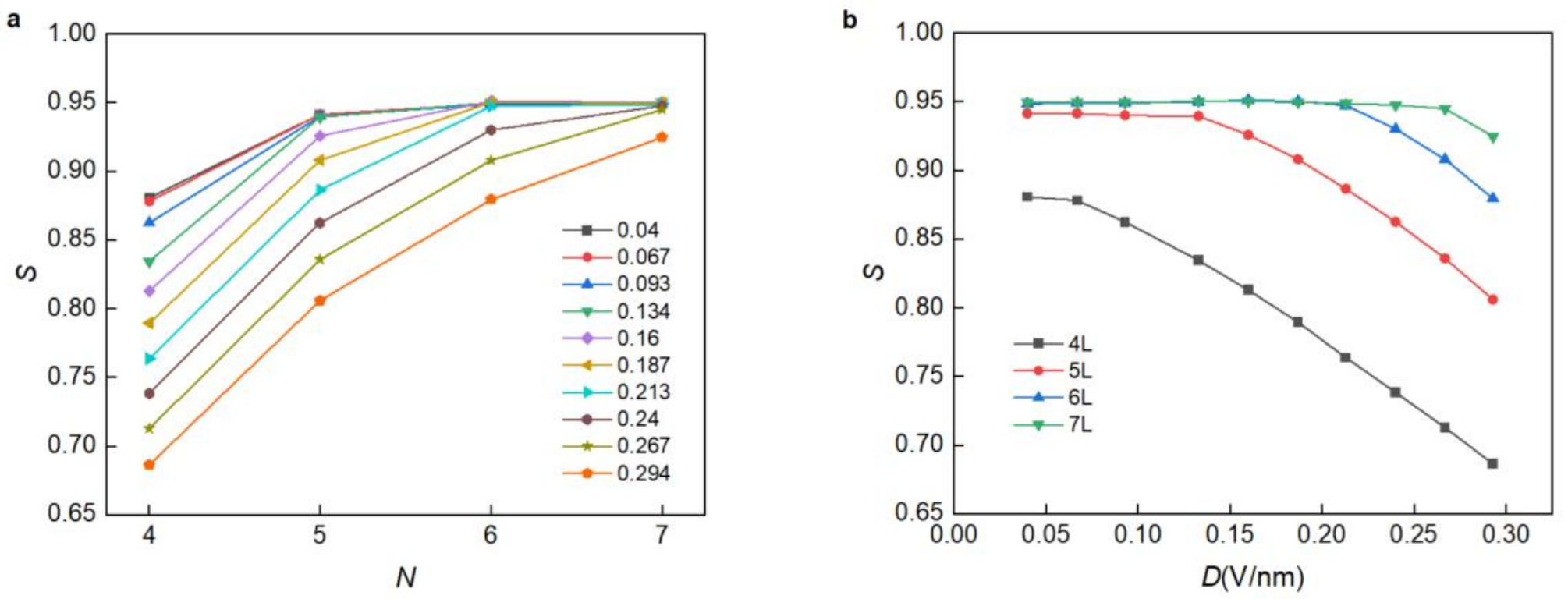

Figure S7. The screening factor $S$ as a function of layer number $N$ at different displacement fields.

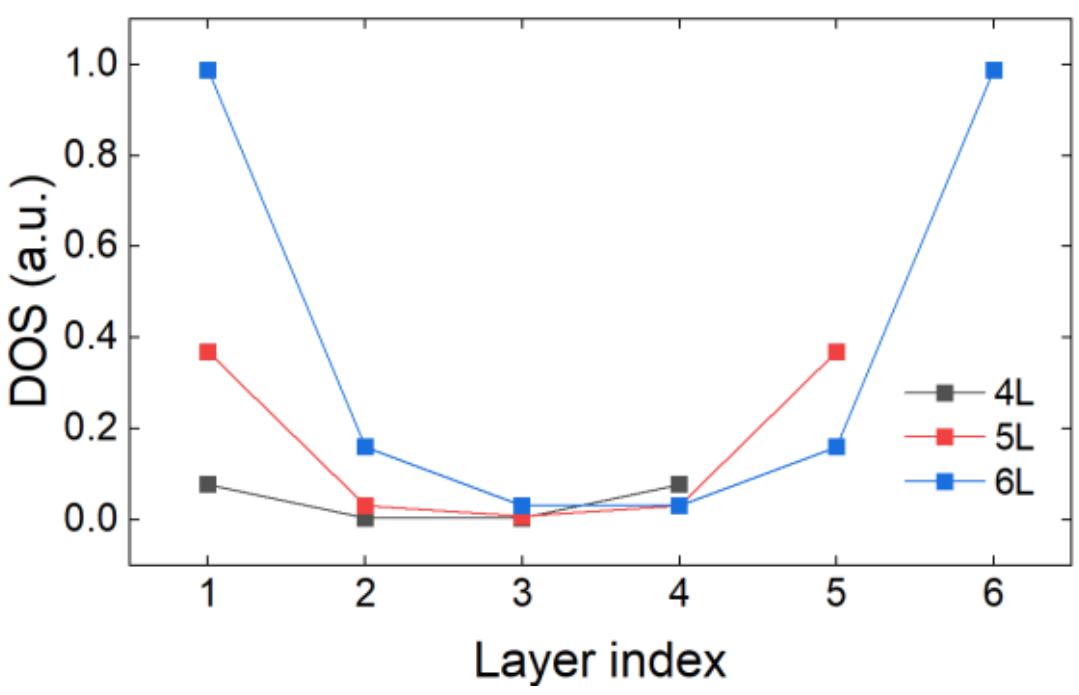

Figure S8. The layer-resolved density of states for *r*-4LG, *r*-5LG, and *r*-6LG at zero external displacement field.

## 7. Resistive peak in the semimetal phase of *r*-6LG

Interestingly, a resistance peak is observed as a function of *D,* in the SM phase. For the understanding of this peak, both carrier densities and mobilities of electrons and holes need to be taken into consideration. Within the SM phase, increasing *D* reduces the band overlap and thus decreases the electron and hole densities, while simultaneously rendering the bands more dispersive and increasing the carrier mobilities. Within a simple Drude picture, the competition between these two effects naturally leads to a maximum in the resistivity at an intermediate displacement field.

## 8. Isospin polarized metals in *r*-6LG

In contrast to the semimetal phases where single-gate dependent resistive features are observed (Fig. S9), spin-and-valley-polarized phase (SVP), spin-polarized phase (SP), and unpolarized phase (UP) are observed in *r*-6LG at relatively higher $n$ and $D$. In these phases, LL features are controlled by both gates (also see Fig. 2(b) in the main text and Fig. S10-S12).

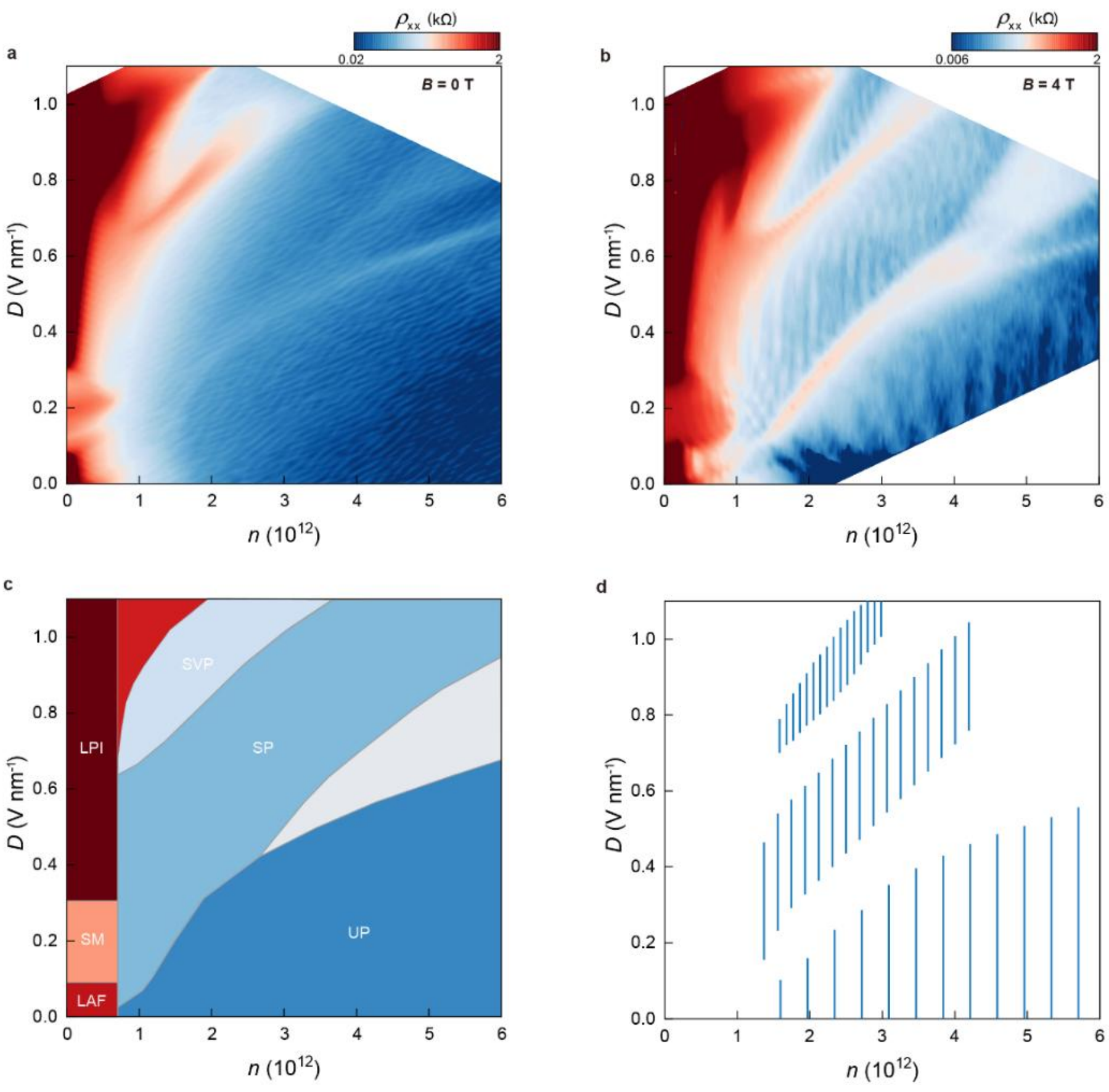

Figure S9. Isospin polarized metals in *r*-6LG. (a and b) longitudinal resistivity $\rho_{xx}$ as a function of displacement field $D$ and carrier density $n$ measured at $B$ = 0 T (a) and 4 T (b). Isospin polarized metal phases separated by resistive are observed. (c) Experimental phase diagram of the broken-symmetry states. "SVP": spin-and-valley-polarized state. "SP": spin-polarized state. "UP": unpolarized state. (d) LLs observed in panel (b), demonstrating different LL degeneracies for three different metal phases.

## 9. Absence of single-gate-controlled behaviors in *r*-4LG and *r*-5LG

In contrast to the *r*-6LG, surface-state-dominated electron transport behaviors are not observed in *r*-4LG and *r*-5LG at different magnetic-field regimes. For example, at $B$ = 1T (Fig. S11(a)), although Landau Levels are not developed yet in *r*-5LG, no single-gate dependent feature appears in the SM phase; At $B$ = 3T (Fig. S10(b)), quantum Hall features defined by total filling factor $n$ start to emerge. Furthermore, at high magnetic field regime, LL filling factor in *r*-4LG and *r*-5LG are strongly dependent on voltages of both gates.

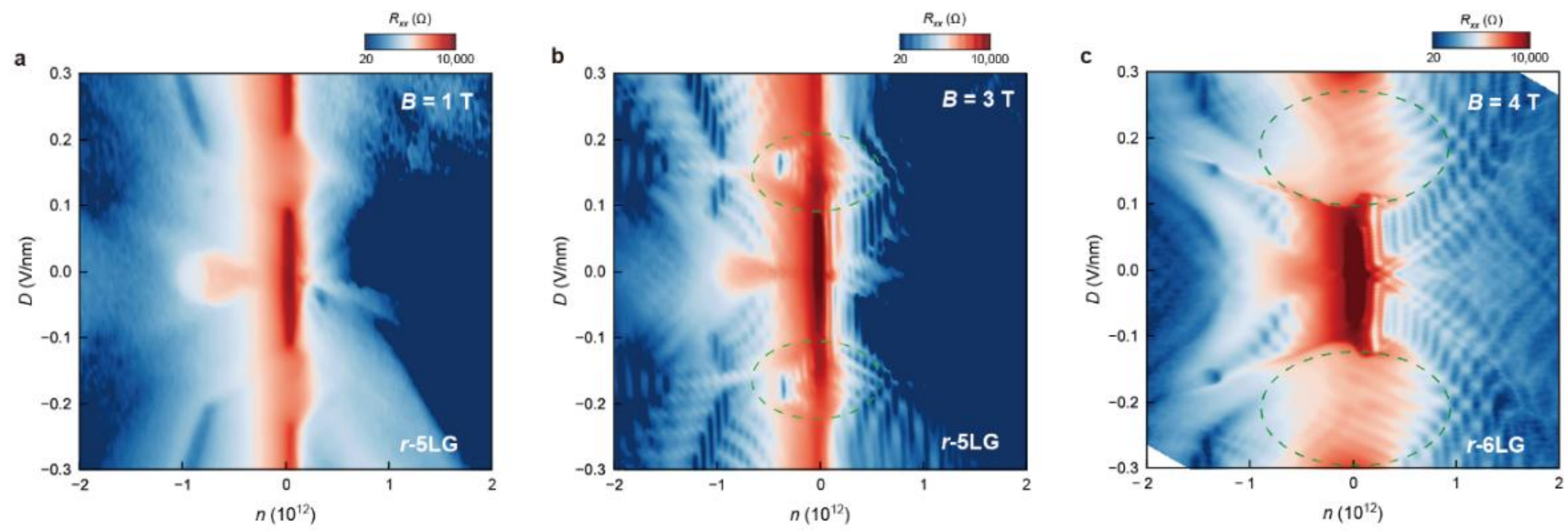

Figure S10. Absence of single-gate-controlled behaviors in *r*-5LG-D3 in contrast to *r*-6LG.

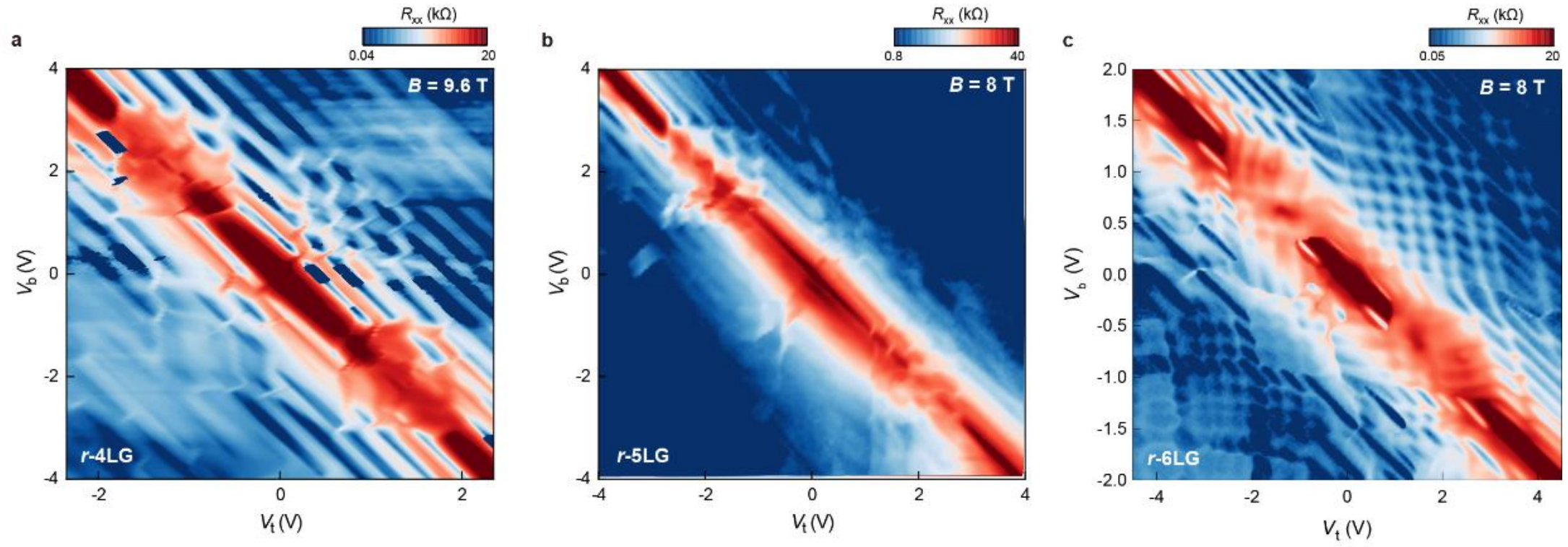

Figure S11. 2D colour plot of longitudinal resistance $R_{xx}$ for *r*-4LG, *r*-5LG and *r*-6LG. In contrast to the *r*-6LG (c), surface-state-dominated electron transport is absent in *r*-4LG (a) and *r*-5LG (b).

## 10. Crossover of the single-gate-controlled and dual-gate-controlled behaviors

At $B$ = 4T and $B$ = 12T, single-gate-controlled LL features and dual-gate-controlled quantum Hall states are observed, respectively. As increasing the magnetic field, a crossover between the two regimes is expected to be observed. Therefore, we present $n$-$D$ colormaps at different magnetic fields for $r$-6LG, as shown in Fig. S12. As increasing out-of-plane external $B$ field, while a sharp critical magnetic field separating single-gate–controlled and dual-gate–controlled behavior cannot be clearly identified, a gradual crossover is evident.

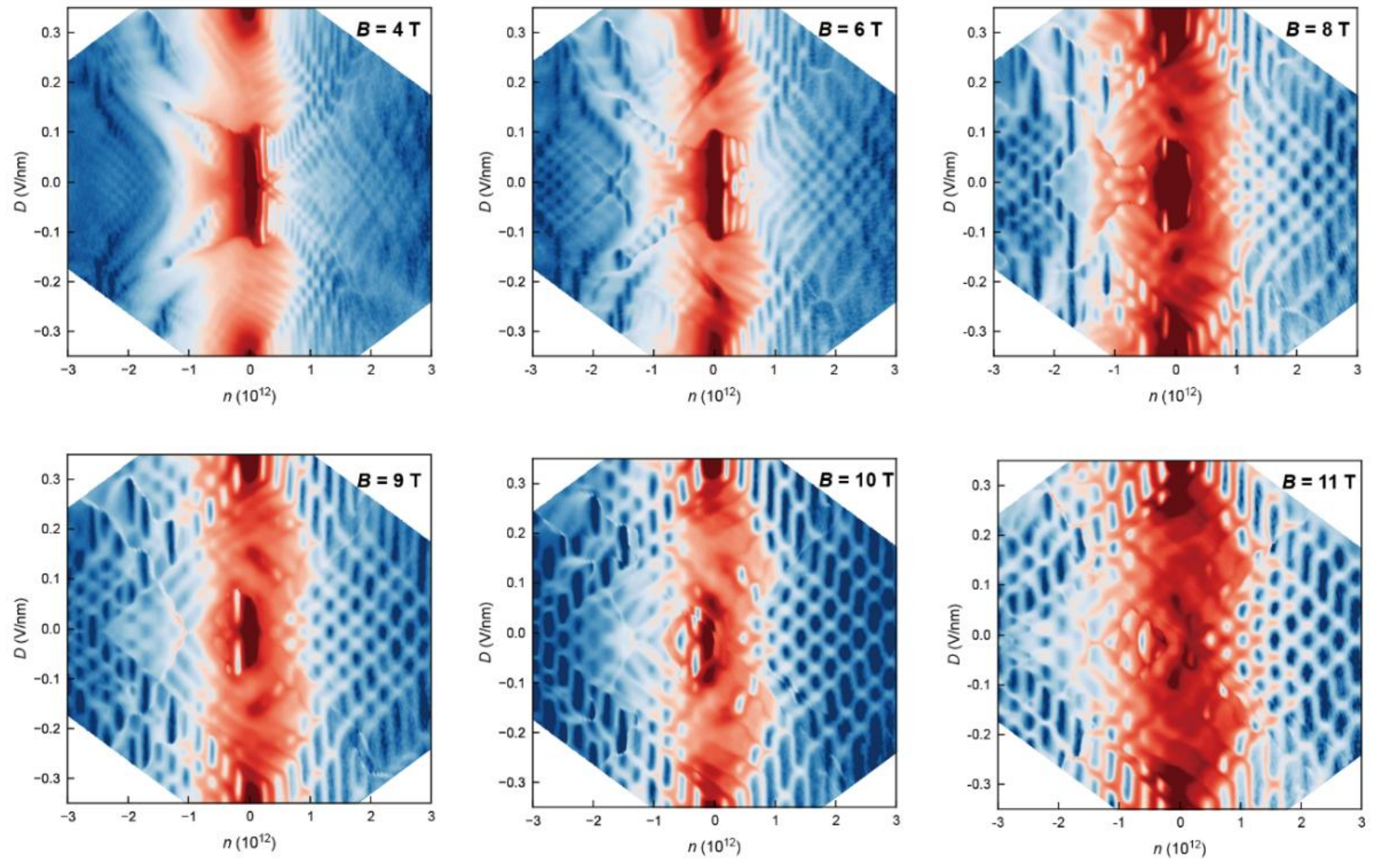

Figure S12. Crossover between the single-gate-controlled to dual-gate-controlled transport behaviors.

## 11. Dissipative and non-dissipative quantum Hall states at $B$ = 12T

Dissipative and non-dissipative quantum Hall states are observed in both $R_{xx}$ (Fig. S13(a)) and $R_{xy}$ data (Fig. S13(b)). Indeed, in the interlayer backscattering regime, $R_{xy}$ does not correspond to the total filling factor. Notably, $R_{xy}$ is negative at $n$ = 0 and even at electron side for the interlayer backscattering patterns, which apparently violate the classical filling factors. For example, along a line cut at total filling factor $\nu = -1$(red dashed line in Fig. S13(b)), the extracted Hall conductance in the range $-0.27 < D < 0.27$ V/nm clearly deviates from the nominal filling factor (Fig. S13(c)). As another example, the "0/3K" state is non-dissipative, whereas its neighboring "−1K′/4K" state exhibits dissipation (Fig. S13(d)).

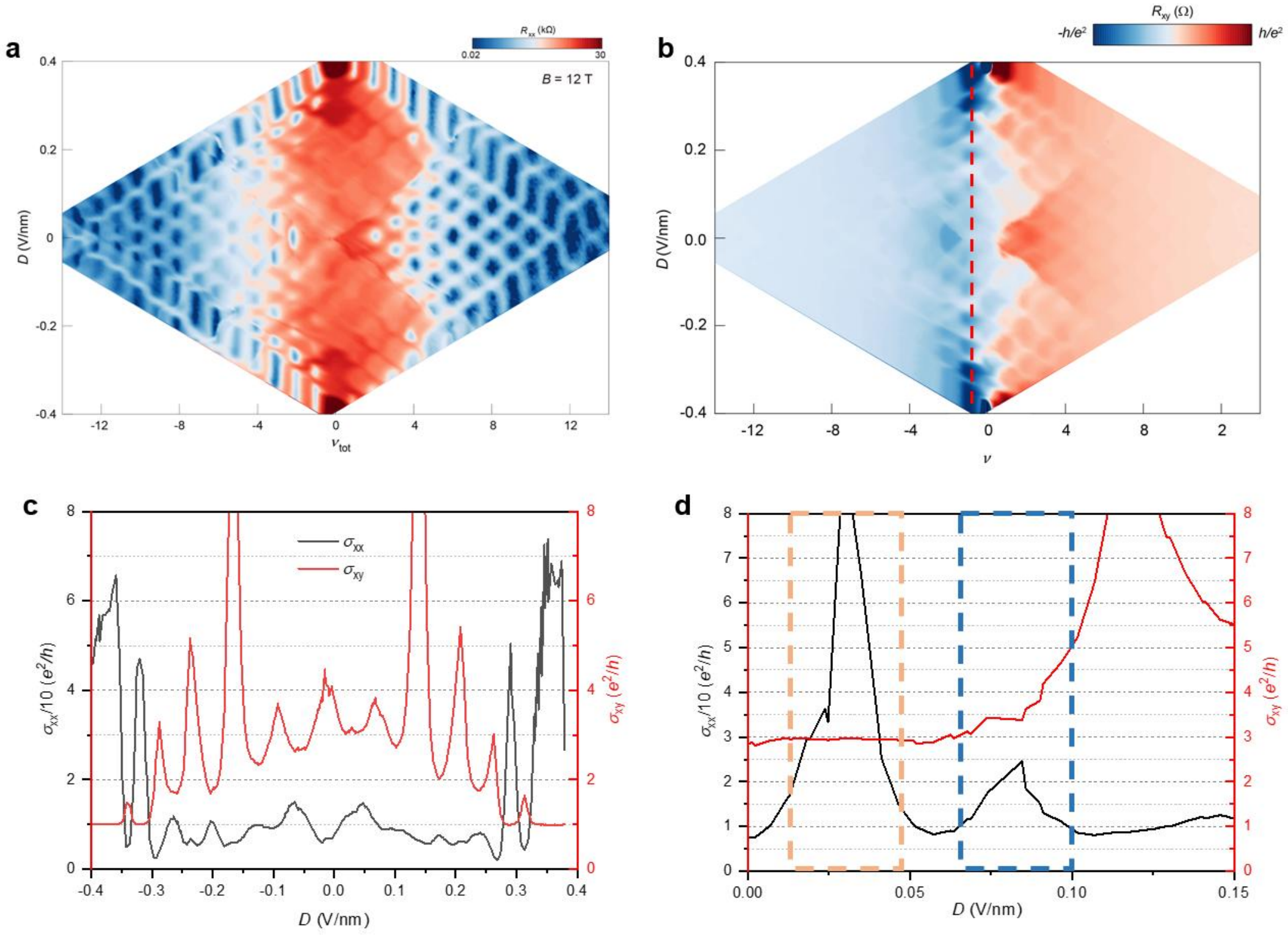

Figure S13. (a) and (b) Color map of $R_{xx}$ and $R_{xy}$ data at $B$ = 12 T. (c) longitudinal and Hall conductivity as a function of $D$ at filling factor $\nu$ = -1 (red dashed line in panel (b)). (d) Comparison of dissipative "-1K'/4K" state and non-dissipative "0/3K" state.

## 12. LL crossing patterns of *r*-6LG and *r*-8LG at high magnetic fields

For a *N* layer rhombohedral graphene, along with the valley and spin degrees of freedom, the zeroth LL contains additional quantum numbers associated with the LL orbital index $n$=0,1,2…$N$-1.[7-9] The zero-energy LL is thus 4$N$-fold degenerate. For *r*-6LG and *r*-8LG, the total filling factors of their lowest LL should be ±12, and ±16, respectively.

At sufficiently high magnetic fields, zeroth-LL degeneracies will be fully lifted, distinctive Landau-level crossing patterns emerge (see panels (a) and (c) of Fig. S14), which are well captured by corresponding schematics (see panels (b) and (d) of Fig. S14). Therefore, in this study, the zeroth-LL crossing patterns observed at high magnetic fields are also used to precisely determine the number of layers of the RG samples.

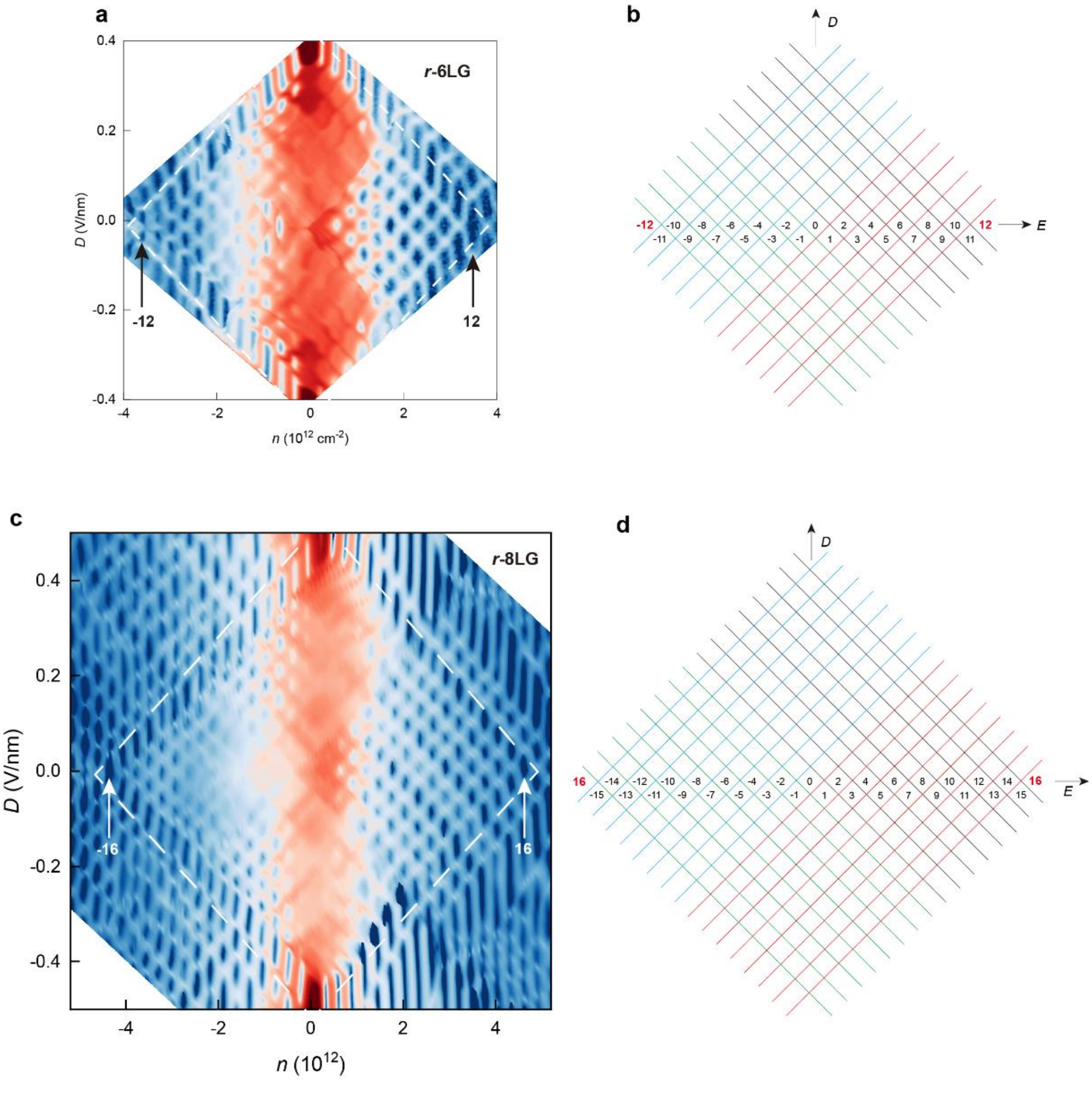

Figure S14. LL crossing patterns in high magnetic fields and corresponding LL crossing diagrams of *r*-6LG (a, b), and *r*-8LG (c,d).